\documentclass[12pt]{article}
\usepackage{graphicx,xcolor}
\usepackage{epsf}
\usepackage{amssymb}
\usepackage{hyperref}

\newcommand{\beq}{\begin{equation}}
\newcommand{\eeq}{\end{equation}}
\newcommand{\bea}{\begin{eqnarray}}
\newcommand{\eea}{\end{eqnarray}}

\begin{document}
\begin{titlepage}
\begin{center}
{\LARGE \bf Mueller-Navelet jets in next-to-leading order

BFKL: theory versus experiment}
\end{center}

\vskip 0.5cm

\centerline{F.~Caporale$^{1\ast}$, D.Yu.~Ivanov$^{2\P}$, B.~Murdaca$^{1\dag}$
and A.~Papa$^{1\ddagger}$}

\vskip .6cm

\centerline{${}^1$ {\sl Dipartimento di Fisica, Universit\`a della Calabria,}}
\centerline{\sl and Istituto Nazionale di Fisica Nucleare, Gruppo collegato di
Cosenza,}
\centerline{\sl Arcavacata di Rende, I-87036 Cosenza, Italy}

\vskip .2cm

\centerline{${}^2$ {\sl Sobolev Institute of Mathematics and Novosibirsk State University,}}
\centerline{\sl 630090 Novosibirsk, Russia}

\vskip 2cm

\begin{abstract}
We study, within QCD collinear factorization and including BFKL
resummation at the next-to-leading order, the production of Mueller-Navelet
jets at LHC with center-of-mass energy of 7 TeV. The adopted jet vertices are
calculated in the approximation of small aperture of the jet cone in the
pseudorapidity-azimuthal angle plane.

We consider several representations of the dijet cross section, differing
only beyond the next-to-leading order, to calculate a few observables
related with this process. We use various methods of optimization to
fix the energy scales entering the perturbative calculation and compare
thereafter our results with the experimental data from the CMS collaboration.
\end{abstract}


$
\begin{array}{ll}
^{\ast}\mbox{{\it e-mail address:}} &
\mbox{francesco.caporale@fis.unical.it}\\
^{\P}\mbox{{\it e-mail address:}} &
\mbox{d-ivanov@math.nsc.ru}\\
^{\dag}\mbox{{\it e-mail address:}} &
\mbox{beatrice.murdaca@fis.unical.it}\\
^{\ddagger}\mbox{{\it e-mail address:}} &
\mbox{alessandro.papa@fis.unical.it}\\
\end{array}
$

\end{titlepage}

\vfill \eject

\section{Introduction}

The investigation of jet production in perturbative QCD is an important
element of phenomenological studies at LHC. Many interesting physical topics
could be studied in such experiments.

In these last years, the inclusive hadroproduction of two jets with large and
similar transverse momenta and a big relative separation in rapidity $Y$, the
so-called Mueller-Navelet jets~\cite{Mueller:1986ey}, has become very popular.
It allows discriminating between BFKL~\cite{BFKL} dynamics of parton-parton
interaction and the standard collinear fixed-order QCD factorization, which 
should work only when $Y$ is not big  enough, $Y\sim 1$. If we compare the 
BFKL dynamics with the fixed-order DGLAP~\cite{DGLAP} calculation, we expect 
a larger cross-section and a reduced azimuthal correlation between the 
detected two forward jets.
If $Y$ is large, the leading terms in a perturbative expansion of the cross 
section (related with forward amplitude) on the coupling $\alpha_s$ are those 
proportional to powers of $\alpha_s Y$, and they are resummed in the BFKL 
series.
At a first, naive analysis Mueller-Navelet jets should manifest an exponential
growth with $Y$, but the hard matrix elements are convoluted via collinear
factorization with the parton distribution functions (PDFs), which damp this
behavior.

Taking into account the effects of the PDFs, it is useful to look for ratios
of distributions. Examples of such ratios are azimuthal angle correlations
between the two measured jets, {\it i.e.} average values of $\cos{(n \phi)}$,
that depend on $Y$ (here $n$ is an integer and $\phi$ is the angle in the
azimuthal plane between the direction of one jet and the opposite direction
of the other jet). Other useful observables are the ratios of two such 
cosines, introduced for the first time in Refs.~\cite{sabioV}.  
We expect a decrease of these observables as $Y$ increases,
due to the larger amount of undetected parton radiation in between the two 
tagged jets.

It is a well known fact that the next-to-leading order (NLO) BFKL corrections
for the $n=0$ conformal spin are with opposite sign with respect to the
leading order (LO) result and large in absolute value. This happens
both to the NLO BFKL kernel~\cite{NLA-kernel}, which enters the integral
equation giving the process-independent BFKL Green's function, and to
process-dependent NLO impact factors (see, {\it e.g.}
Ref.~\cite{Ivanov2006}, for the case of the vector meson photoproduction).
The impact factor needed for the BFKL description of the Mueller-Navelet
jet production, the so-called forward jet vertex~\cite{IFjet,SCA}, is of no
exception. For this reason it is strictly necessary to optimize the
amplitude by (i) including some pieces of the (unknown) next-to-NLO
corrections and/or (ii) suitably choosing the values of the energy and
renormalization scales, which, though being arbitrary within the NLO, can
have a sizeable numerical impact through subleading terms.
A remarkable example of the former approach is the so-called \emph{collinear
improvement}~\cite{collinear}, based on the inclusion of terms generated by
renormalization group (RG), or collinear, analysis, leading to more
convergent kernels. As for the latter approach, the most common ways to
optimize the choice of the energy and renormalization scales are those
inspired by the \emph{principle of minimum sensitivity} (PMS)~\cite{PMS},
the \emph{fast apparent convergence} (FAC)~\cite{FAC} and the
\emph{Brodsky-LePage-McKenzie method} (BLM)~\cite{BLM}.

In an ideal situation, the use of one or the other optimization procedure
should not change much the prediction for any of the observables related
with a given process. In practice, this may well not be the case. Then, it
becomes fundamental to identify those observables, if any, which show no
or small sensitivity to the change of optimization procedure. Otherwise, the
preference to one optimization procedure should be assigned by evaluating the
agreement with the experimental data in a certain setup and, thereafter,
assumed to apply also in other setups.

The study of Mueller-Navelet jet production process at LHC is, in this respect,
a paradigmatic case. The first, pioneer paper devoted to the study
of this process within full NLO BFKL~\cite{Colferai2010} used kinematical 
({\it i.e.} non-optimized) energy scales and considered also as an option
the case of an RG-improved kernel. Here the predictions for differential 
cross section and several azimuthal correlations at the design LHC 
center-of-mass energy of 14 TeV were built.  Later, a similar analysis was 
redone~\cite{Caporale2013}, using the standard ({\it i.e.} non-RG-improved) 
kernel, but energy scales optimized according to the PMS method. Besides, 
in~\cite{Caporale2013} the analytic expressions for jet vertices derived 
in a small-cone approximation~\cite{SCA} were used.
The small-cone approximation allows to simplify the numerical analysis and
is an adequate tool since typically the difference between it and the exact 
jet definition is much smaller than other theoretical uncertainties inherent 
to the BFKL approach.
A third paper~\cite{Salas2013} followed the same approach of
Ref.~\cite{Caporale2013}, but adopted an RG-improved kernel and observed
a tendency of optimal values of the energy scales towards ``naturalness''.

The appearance of the first CMS data at a center-of-mass energy of
7 TeV~\cite{CMS} triggered the theoretical analysis in the same kinematical
setup, which showed that the use of a RG-improved kernel with non-optimized
energy scales does not lead to agreement with the
experiment~\cite{Ducloue2013}, but a nice agreement is found at the larger
values of $Y$ when BLM-optimal energy scales are used
instead~\cite{Ducloue2014}, both in pure BFKL and RG-improved calculations.
Recently some effects subleading to the BFKL 
approach, dubbed as ``violation of the energy-momentum conservation'', 
were studied in the context of the Mueller-Navelet jet production 
process~\cite{Ducloue:2014koa}.

The aim of the present paper is to supplement the nice results
achieved in Refs.~\cite{Ducloue2013,Ducloue2014} with some further information.
In particular, we will try to answer, at least partially, to the following
questions:

- are there observables weakly sensitive (or insensitive at all) to the
optimization procedure?

- do other optimization schemes, such as PMS and FAC, reproduce the
CMS experimental data as well as BLM, if necessary by modifying the
amplitude with the inclusion of some of the unknown next-to-NLO corrections?

- does the BLM method reproduce experimental data also for the total 
Mueller-Navelet cross section as it does for azimuthal correlations?

The paper is organized as follows: in the next Section we will give the
kinematics and the basic formulae for the Mueller-Navelet jet process
cross section, present the different, NLO-equivalent representations of the
amplitude adopted in this work and briefly recall the PMS, FAC and BLM
optimization methods; in Section~\ref{results} we will present our results;
finally, in Section~\ref{summary} we will draw our conclusions and discuss 
some issues which we believe to be important in confronting the theoretical 
predictions with experimental data.

\section{The Mueller-Navelet jet process}
\label{intro}

We consider the production of Mueller-Navelet jets~\cite{Mueller:1986ey} in
proton-proton collisions
\begin{eqnarray}
\label{process}
p(p_1) + p(p_2) \to {\rm jet}(k_{J_1}) + {\rm jet}(k_{J_2})+ X \;,
\end{eqnarray}
where the two jets are characterized by high transverse momenta,
$\vec k_{J_1}^2\sim \vec k_{J_2}^2\gg \Lambda_{QCD}^2$ and large separation
in rapidity; $p_1$ and $p_2$ are taken as Sudakov vectors satisfying
$p_1^2=p_2^2=0$ and $2\left( p_1 p_2\right)=s$.

\begin{figure}[p]
\centering
  \includegraphics[scale=0.7]{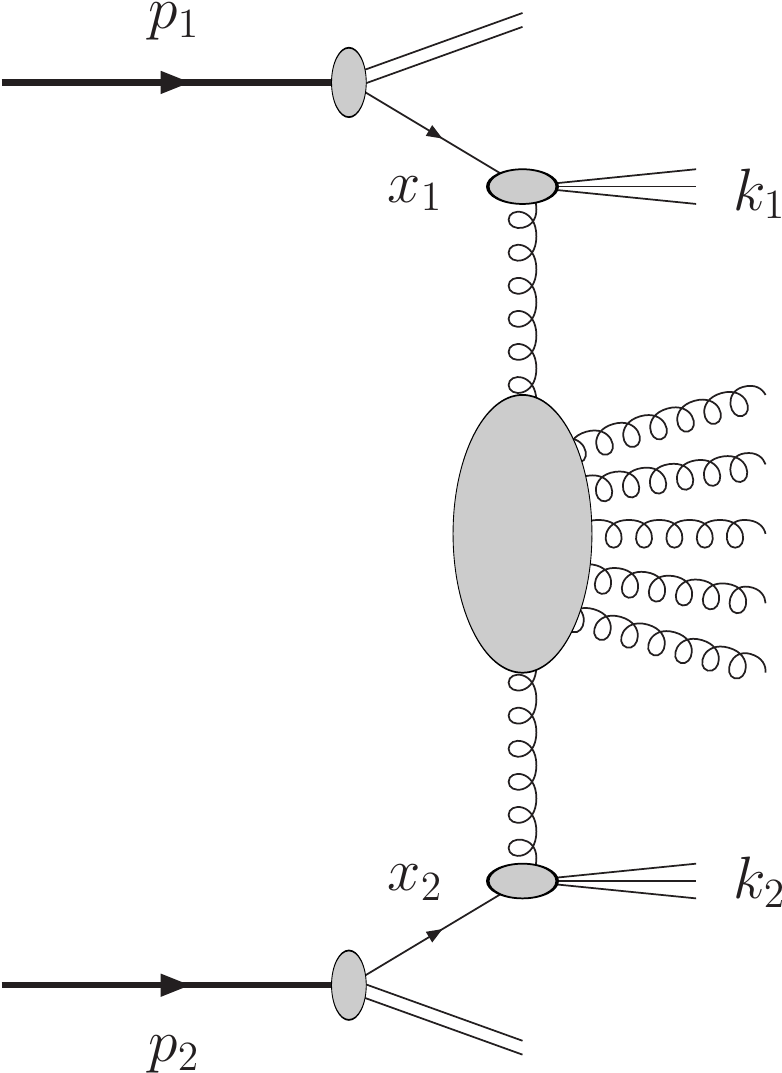}
\caption{Mueller-Navelet jet production process.}
\label{fig:MN}
\end{figure}

In QCD collinear factorization the cross section of the process~(\ref{process})
reads
\beq
\frac{d\sigma}{dx_{J_1}dx_{J_2}d^2k_{J_1}d^2k_{J_2}}
=\sum_{i,j=q,{\bar q},g}\int_0^1 dx_1 \int_0^1 dx_2\ f_i\left(x_1,\mu_F\right)
\ f_j\left(x_2,\mu_F\right)\frac{d{\hat\sigma}_{i,j}\left(x_1x_2s,\mu_F\right)}
{dx_{J_1}dx_{J_2}d^2k_{J_1}d^2k_{J_2}}\;,
\eeq
where the $i, j$ indices specify the parton types (quarks $q = u, d, s, c, b$;
antiquarks $\bar q = \bar u, \bar d, \bar s, \bar c, \bar b$; or gluon $g$),
$f_i\left(x, \mu_F \right)$ denotes the initial proton PDFs; $x_{1,2}$ are
the longitudinal fractions of the partons involved in the hard subprocess,
while $x_{J_{1,2}}$ are the jet longitudinal fractions; $\mu_F$ is the
factorization scale; $d\hat\sigma_{i,j}\left(x_1x_2s, \mu_F \right)$ is
the partonic cross section for the production of jets and
$x_1x_2s\equiv\hat s$ is the squared center-of-mass energy of the
parton-parton collision subprocess (see Fig.~\ref{fig:MN}).

In the BFKL approach~\cite{BFKL}, the cross section of the hard subprocess
can be written as (see Ref.~\cite{Caporale2013} for the details of the
derivation)
\beq
\frac{d\sigma}
{dy_{J_1}dy_{J_2}\, d|\vec k_{J_1}| \, d|\vec k_{J_2}|
d\phi_{J_1} d\phi_{J_2}}
=\frac{1}{(2\pi)^2}\left[{\cal C}_0+\sum_{n=1}^\infty  2\cos (n\phi )\,
{\cal C}_n\right]\, ,
\eeq
where $\phi=\phi_{J_1}-\phi_{J_2}-\pi$ and the cross section ${\cal C}_0$
and the other coefficients ${\cal C}_n$ are given by
\beq
\label{Cm}
{\cal C}_n \equiv \int_0^{2\pi}d\phi_{J_1}\int_0^{2\pi}d\phi_{J_2}\,
\cos[n(\phi_{J_1}-\phi_{J_2}-\pi)] \,
\frac{d\sigma}{dy_{J_1}dy_{J_2}\, d|\vec k_{J_1}| \, d|\vec k_{J_2}|
d\phi_{J_1} d\phi_{J_2}}\;
\eeq
\[
= \frac{x_{J_1} x_{J_2}}{|\vec k_{J_1}| |\vec k_{J_2}|}
\int_{-\infty}^{+\infty} d\nu \, \left(\frac{x_{J_1} x_{J_2} s}{s_0}
\right)^{\bar \alpha_s(\mu_R)\chi(n,\nu)}
\]
\[
\times \alpha_s^2(\mu_R) c_1(n,\nu,|\vec k_{J_1}|, x_{J_1})
c_2(n,\nu,|\vec k_{J_2}|,x_{J_2})\,
\]
\[
\times \left[1
+\alpha_s(\mu_R)\left(\frac{c_1^{(1)}(n,\nu,|\vec k_{J_1}|,
x_{J_1})}{c_1(n,\nu,|\vec k_{J_1}|, x_{J_1})}
+\frac{c_2^{(1)}(n,\nu,|\vec k_{J_2}|, x_{J_2})}{c_2(n,\nu,|\vec k_{J_2}|,
x_{J_2})}\right)
\right.
\]
\[
\left.
+\bar \alpha_s^2(\mu_R) \ln\left(\frac{x_{J_1}x_{J_2} s}{s_0}\right)
\left(\bar \chi(n,\nu)
+ \frac{\beta_0}{8C_A}\chi(n,\nu)
\left(-\chi(n,\nu) + \frac{10}{3} + \ln\frac{\mu_R^4}
{\vec k_{J_1}^2 \vec k_{J_2}^2}\right)\right)\right] \;.
\]
Here $\bar \alpha_s(\mu_R) \equiv \alpha_s(\mu_R) N_c/\pi$, with
$N_c$ the number of colors,
\beq
\beta_0=\frac{11}{3} N_c - \frac{2}{3}n_f
\eeq
is the first coefficient of the QCD $\beta$-function,
\beq
\chi\left(n,\nu\right)=2\psi\left( 1\right)-\psi\left(\frac{n}{2}
+\frac{1}{2}+i\nu \right)-\psi\left(\frac{n}{2}+\frac{1}{2}-i\nu \right)
\eeq
is the LO BFKL characteristic function,
\beq
\label{c1}
c_1(n,\nu,|\vec k|,x)=2\sqrt{\frac{C_F}{C_A}}
(\vec k^{\,2})^{i\nu-1/2}\,\left(\frac{C_A}{C_F}f_g(x,\mu_F)
+\sum_{a=q,\bar q}f_q(x,\mu_F)\right)
\eeq
and
\beq
\label{c2}
c_2(n,\nu,|\vec k|,x)=\biggl[c_1(n,\nu,|\vec k|,x) \biggr]^* \;,
\eeq
are the LO jet vertices in the $\nu$-representation. The remaining objects
are related with the NLO corrections of the BFKL kernel ($\bar \chi(n,\nu)$)
and of the jet vertices in the small-cone approximation
($c_{1,2}^{(1)}(n,\nu,|\vec k_{J_2}|, x_{J_2})$)
in the $\nu$-representation. Their expressions are given in Eqs.~(23), (36)
and~(37) of Ref.~\cite{Caporale2013}.

The representation~(\ref{Cm}) is valid both in the leading logarithm
approximation (LLA), which means resummation of leading energy logarithms,
all terms $\left( \alpha_s\ln\left(s\right)\right)^n$, and in the
next-to-leading approximation (NLA), which means resummation of all
terms $\alpha_s\left(\alpha_s\ln\left(s\right)\right)^n$. The scale $s_0$
is artificial. It is introduced in the BFKL approach at the time to perform
the Mellin transform from the $s$-space to the complex angular momentum plane
and cancels in the full expression, up to terms beyond the NLA.

Eq.~(\ref{Cm}) represents just one of infinitely many representations of
the coefficients ${\cal C}_n$. One can consider alternative representations,
aiming at catching some of the unknown next-to-NLA corrections. Introducing
for the sake of brevity the definitions
\[
Y=\ln\frac{x_{J_1}x_{J_2}s}{|\vec k_{J_1}||\vec k_{J_2}|}\;,
\;\;\;\;\;
Y_0=\ln\frac{s_0}{|\vec k_{J_1}||\vec k_{J_2}|}\;,
\]
the representations we will use in this work are the following:
\begin{itemize}
\item the so-called \emph{exponentiated} representation,
\begin{eqnarray}
\nonumber
{\cal C}_n^{\rm exp}&=&\frac{x_{J_1} x_{J_2}}{|\vec k_{J_1}| |\vec k_{J_2}|}\int_{-\infty}^{+\infty}d\nu\ e^{(Y-Y_0)
\bigl[\bar \alpha_s\left(\mu_R\right)\chi\left(n,\nu\right)
+\bar \alpha_s^2\left(\mu_R\right)\left(\chi^{\left(1\right)}\left(n,\nu\right)+ \frac{\beta_0}{8C_A}\chi\left(n,\nu\right)\log\frac{\mu_R^4}{\vec k_{J_1}^2\vec k_{J_2}^2}
\right)\bigr]}
\alpha_s^2\left(\mu_R\right)
\\ \label{pure}
&\times&c_1\left(n,\nu\right)
c_2\left(n,\nu\right)\left[ 1+ \alpha_s\left(\mu_R\right)
\left( \frac{c_1^{\left(1\right)}\left(n,\nu\right)}{c_1\left(n,\nu\right)}
+\frac{c_2^{\left(1\right)}\left(n,\nu\right)}{c_2\left(n,\nu\right)}\right)
\right]\;,
\end{eqnarray}
where the dependence on $|\vec k_{J_i}|$ and $x_{J_i}$ in $c_{1,2}^{(1)}$
has been omitted for simplicity and
\[
\chi^{\left(1\right)}\left(n,\nu\right)=\bar\chi\left(n,\nu\right)
+\frac{\beta_0}{8C_A}\chi\left(n,\nu\right)\left(-\chi\left(n,\nu\right)
+\frac{10}{3}\right)\;,
\]
with $\bar \chi(n,\nu)$ given by Eq.~(23) in Ref.~\cite{Caporale2013}.

\item the exponentiated representation with an extra, irrelevant in the
NLA term, given by the product of the NLO corrections of the two jet vertices,
\begin{eqnarray}
\nonumber
{\cal C}_n^{\rm sq}&=&\frac{x_{J_1} x_{J_2}}{|\vec k_{J_1}| |\vec k_{J_2}|}\int_{-\infty}^{+\infty}d\nu\ e^{(Y-Y_0)
\bigl[\bar \alpha_s\left(\mu_R\right)\chi\left(n,\nu\right)
+\bar \alpha_s^2\left(\mu_R\right)\left(\chi^{\left(1\right)}\left(n,\nu\right)+ \frac{\beta_0}{8C_A}\chi\left(n,\nu\right)\log\frac{\mu_R^4}{\vec k_{J_1}^2\vec k_{J_2}^2}
\right)\bigr]}
\alpha_s^2\left(\mu_R\right)
\\
\nonumber
&\times&c_1\left(n,\nu\right)
c_2\left(n,\nu\right)\left[ 1+ \alpha_s\left(\mu_R\right)
\left( \frac{c_1^{\left(1\right)}\left(n,\nu\right)}{c_1\left(n,\nu\right)}
+\frac{c_2^{\left(1\right)}\left(n,\nu\right)}{c_2\left(n,\nu\right)}\right)\right.
\\
\label{square}
&+&\left.  \alpha_s^2\left(\mu_R\right)
\left( \frac{c_1^{\left(1\right)}\left(n,\nu\right)}{c_1\left(n,\nu\right)}
\frac{c_2^{\left(1\right)}\left(n,\nu\right)}{c_2\left(n,\nu\right)}\right)
\right]\;,
\end{eqnarray}

\item the exponentiated representation with an RG-improved kernel,
\begin{eqnarray}
\nonumber
{\cal C}_n^{\rm RG}&=&\frac{x_{J_1} x_{J_2}}{|\vec k_{J_1}| |\vec k_{J_2}|}\int_{-\infty}^{+\infty}d\nu\ e^{(Y-Y_0)
\bigl[\bar \alpha_s\left(\mu_R\right)\chi\left(n,\nu\right)
+\bar \alpha_s^2\left(\mu_R\right)\left(\chi^{\left(1\right)}\left(n,\nu\right)+ \frac{\beta_0}{8C_A}\chi\left(n,\nu\right)\log\frac{\mu_R^4}{\vec k_{J_1}^2\vec k_{J_2}^2}
\right)+\chi_{RG}\left(n,\nu\right)\bigr]}
\\
 \label{ci}
&\times&\alpha_s^2\left(\mu_R\right)c_1\left(n,\nu\right)
c_2\left(n,\nu\right)\left[ 1+ \alpha_s\left(\mu_R\right)
\left( \frac{c_1^{\left(1\right)}\left(n,\nu\right)}{c_1\left(n,\nu\right)}
+\frac{c_2^{\left(1\right)}\left(n,\nu\right)}{c_2\left(n,\nu\right)}\right)
\right]\;,
\end{eqnarray}
where $\chi^{(1)}_{\rm RG}(n,\nu)$ is given in Eqs.~(13)-(15) of
Ref.~\cite{Salas2013};

\item a combination of the previous two representations,
\begin{eqnarray}
\nonumber
{\cal C}_n^{\rm RG+sq}&=& \int_{-\infty}^{+\infty}d\nu\ e^{(Y-Y_0)
\bigl[\bar \alpha_s\left(\mu_R\right)\chi\left(n,\nu\right)
+\bar \alpha_s^2\left(\mu_R\right)\left(\chi^{\left(1\right)}\left(n,\nu\right)+ \frac{\beta_0}{8C_A}\chi\left(n,\nu\right)\log\frac{\mu_R^4}{\vec k_{J_1}^2\vec k_{J_2}^2}
\right)+\chi_{RG}\left(n,\nu\right)\bigr]}
\\
\nonumber
&\times&\frac{x_{J_1} x_{J_2}}{|\vec k_{J_1}| |\vec k_{J_2}|}\alpha_s^2\left(\mu_R\right)c_1\left(n,\nu\right)
c_2\left(n,\nu\right)\left[ 1+ \alpha_s\left(\mu_R\right)
\left( \frac{c_1^{\left(1\right)}\left(n,\nu\right)}{c_1\left(n,\nu\right)}
+\frac{c_2^{\left(1\right)}\left(n,\nu\right)}{c_2\left(n,\nu\right)}\right)
\right.
\\
\label{cis}
&+&\left. \alpha_s^2\left(\mu_R\right)
\left( \frac{c_1^{\left(1\right)}\left(n,\nu\right)}{c_1\left(n,\nu\right)}
\frac{c_2^{\left(1\right)}\left(n,\nu\right)}{c_2\left(n,\nu\right)}\right)
\right]\;.
\end{eqnarray}

\end{itemize}

\section{Numerical results}
\label{results}

In this Section we present our results for the dependence on
$Y=y_{J_1}-y_{J_2}$ of the coefficients ${\cal C}_n$ and of their
ratios ${\cal R}_{nm}\equiv{\cal C}_n/{\cal C}_m$. Among them, the ratios of
the form $R_{n0}$ have a simple physical interpretation, being the azimuthal
correlations $\langle \cos(n\phi)\rangle$.

In order to match the kinematical cuts used by the CMS collaboration, we will
consider the \emph{integrated coefficients} given by
\beq
\label{Cm_int}
C_n=\int_{y_{1,\rm min}}^{y_{1,\rm max}}dy_1
\int_{y_{2,\rm min}}^{y_{2,\rm max}}dy_2\int_{k_{J_1,\rm min}}^{\infty}dk_{J_1}
\int_{k_{J_2,\rm min}}^{\infty}dk_{J_2} \delta\left(y_1-y_2-Y\right){\cal C}_n
\left(y_{J_1},y_{J_2},k_{J_1},k_{J_2} \right)\;,
\eeq
with $y_{1,\rm min}=y_{2,\rm min}=-4.7$, $y_{1,\rm max}=y_{2,\rm max}=4.7$,
$k_{J_1,\rm min}=k_{J_2,\rm min}=35$ GeV, and their ratios $R_{nm}\equiv
C_n/C_m$. We fix the jet cone size at the
value $R=0.5$ and the center-of-mass energy at $\sqrt s=7$ TeV.
We use the PDF set MSTW2008nlo~\cite{PDF} and the two-loop
running coupling with $\alpha_s\left(M_Z\right)=0.11707$.

As discussed in the Introduction, to improve the stability of the perturbative
series, which is particularly relevant in the BFKL framework, several methods
have been devised for the optimal choice of the several energy scales entering
the above expressions. We will use the following:
\begin{itemize}
\item \emph{principle of minimal sensitivity} (PMS)~\cite{PMS},
\item \emph{fast apparent convergence} (FAC)~\cite{FAC},
\item \emph{Brodsky-LePage-McKenzie} (BLM) method~\cite{BLM}.
\end{itemize}

\subsection{PMS}

We used an adaptation of the standard PMS method, as usual in our works,
valid when more than one energy scale is present. The optimal choices for
$\mu_R$ and $s_0$ are those values for which the physical observable under
exam exhibits the minimal sensitivity under variation of both these scales.

We applied the method to the four representations given in
Eqs.~(\ref{pure})-(\ref{cis}). As for the optimal choice of the third scale,
the factorization scale $\mu_F$, we considered the following two options:

(i) let $\mu_F$ follow the same fate of the renormalization scale $\mu_R$,

(ii) fix $\mu_F$ at $|\vec k_{J_1}|$ in the vertex of the jet~1 and
at $|\vec k_{J_2}|$ in the vertex of the jet~2.

This leads to consider the eight following possibilities:

\begin{tabular}{cll}
\vspace{0.3cm}
NLA$_1$, \ \ & $[{\cal C}_n^{\rm exp}]_{\mu_F=\mu_R}$
& (Eq.~(\ref{pure}) + option~(i); dark green in Figs.~\ref{fig:PMS})\\
\vspace{0.3cm}
NLA$_2$, \ \ & $[{\cal C}_n^{\rm exp}]_{\mu_F=k_{J_i}}$
& (Eq.~(\ref{pure}) + option~(ii); green in Figs.~\ref{fig:PMS})\\
\vspace{0.3cm}
NLA$_3$, \ \ & $[{\cal C}_n^{\rm sq}]_{\mu_F=\mu_R}$
& (Eq.~(\ref{square})+ option~(i); violet in Figs.~\ref{fig:PMS})\\
\vspace{0.3cm}
NLA$_4$, \ \ & $[{\cal C}_n^{\rm sq}]_{\mu_F=k_{J_i}}$
& (Eq.~(\ref{square}) + option~(ii); magenta in Figs.~\ref{fig:PMS})\\
\vspace{0.3cm}
NLA$_5$, \ \ & $[{\cal C}_n^{\rm RG}]_{\mu_F=\mu_R}$
& (Eq.~(\ref{ci}) + option~(i); blue in Figs.~\ref{fig:PMS})\\
\vspace{0.3cm}
NLA$_6$, \ \ & $[{\cal C}_n^{\rm RG}]_{\mu_F=k_{J_i}}$
& (Eq.~(\ref{ci})+ option~(ii); cyan in Figs.~\ref{fig:PMS})\\
\vspace{0.3cm}
NLA$_7$, \ \ & $[{\cal C}_n^{\rm RG+sq}]_{\mu_F=\mu_R}$
& (Eq.~(\ref{cis}) + option~(i); black in Figs.~\ref{fig:PMS})\\
\vspace{0.3cm}
NLA$_8$, \ \ & $[{\cal C}_n^{\rm RG+sq}]_{\mu_F=k_{J_i}}$
& (Eq.~(\ref{cis}) + option~(ii); gray in Figs.~\ref{fig:PMS})\\
\end{tabular}

Following Refs.~\cite{Ivanov2006,Caporale2013,Salas2013}, in our search of
the optimal values for the $Y_0$ and $\mu_R$, we considered integer values
for $Y_0$ in the range $0\div 6$ and values for $\mu_R$ given by multiples
of $\sqrt{|\vec k_{J_1}||\vec k_{J_2}|}$,
\beq
\mu_R=n_R\sqrt{|\vec k_{J_1}||\vec k_{J_2}|}\;,
\eeq
with the integer $n_R$ in the range $1\div 9$.

We looked for stationary points of the coefficient $C_n$ in the $Y_0-n_R$
plane, then the ratios $C_n/C_m$ were obtained indirectly by using the
optimal results for the coefficients $C_n$ and $C_m$. In particular,
following Ref.~\cite{CMS}, we studied the ratios $R_{10}$, $R_{20}$, $R_{30}$,
$R_{21}$ and $R_{32}$. We carried out this analysis for all the
representations NLA$_i$, $i$=1,...,8, listed above. Results are reported in
Tables~\ref{tab:C1C0_PMS}-\ref{tab:C3C2_PMS} and in Figs.~\ref{fig:PMS}.
For the sake of brevity, we do not show in these Tables the optimal values
of $Y_0-n_R$, but simply say that they are quite sparse in the given
intervals, with more recurrent values for $Y_0$ in the range $2\div 5$
and for $n_R$ in the range $2\div 6$.

We can see that the theoretical predictions overshoot data at all values
of $Y$ in the cases of $C_1/C_0$ and $C_2/C_0$ and at the smaller $Y$'s
for $C_3/C_0$, while there is a agreement, at least for some of the
eight options, for the ratios $C_2/C_1$ and $C_3/C_2$.

\subsection{FAC}

This method consists in fixing the renormalization scale $\mu_R$ at
the value for which the highest-order correction term is exactly zero. In
our case, the application of this method requires an adaptation, since
there is a second energy parameter to take care of, $Y_0$.

We applied it to the representation labeled by NLA$_1$ and, for each $Y_0$
in a finite set of integer and half-integer values in the range 0-6, we found
the value of $\mu_R$ such that the highest-order correction term of a
certain coefficient $C_n$ is exactly zero. Then, a stationary point was
searched for varying $Y_0$ in the given set.

This method in general did not allow to find clear regions of stability.
Nevertheless, we report some of our results in Table~\ref{tab:FAC}, for
the sake of comparison with the other methods.

\subsection{BLM}

This method consists in choosing the scale $\mu_R$ such that it makes vanish
completely the $\beta_0$-dependence of a given observable.

Also in this case we considered only the representation labeled by NLA$_1$,
{\it i.e.} the exponential representation with $\mu_F=\mu_R$. We implemented
the BLM procedure in a slightly different way from Ref.~\cite{Ducloue2014}.
As a matter of fact, we realized that a clear-cut way to implement this
procedure in the present case is not obviously found. We rather implemented
two variants of the BLM method, dubbed $(a)$ and $(b)$, and give here
all the relevant formulae, but refer to a separate publication for
details~\cite{BLMpaper}.

The variant $(a)$ is given by
\begin{eqnarray}
\nonumber
{\cal C}_n^{\rm exp-BLM_a}&=& \frac{x_{J_1}x_{J_2}}{|\vec k_{J_1}|
|\vec k_{J_2}|}\int_{-\infty}^{+\infty}d\nu
\ e^{(Y-Y_0)\left[\bar \alpha_s\left(\mu_R\right)\chi\left(n,\nu\right)
+\bar \alpha_s^2\left(\mu_R\right)
\left( \bar \chi\left(n,\nu\right)-\frac{T^{\beta}}{C_A}\chi\left(n,\nu\right)
-\frac{\beta_0}{8C_A}\chi^2\left(n,\nu\right)\right)\right]}\alpha_s^2
\left(\mu_R\right)
\\
\label{casea}
&\times&c_1\left(n,\nu\right)
c_2\left(n,\nu\right)\left[ 1-\frac{2}{\pi}\alpha_s\left(\mu_R\right)T^{\beta}
+ \alpha_s\left(\mu_R\right) \left( \frac{\bar c_1^{\left(1\right)}
\left(n,\nu\right)}{c_1\left(n,\nu\right)}+\frac{\bar c_2^{\left(1\right)}
\left(n,\nu\right)}{c_2\left(n,\nu\right)}\right)
\right]\;,
\end{eqnarray}
with $\mu_R$ fixed at the value
\beq
(\mu_R^{\rm BLM})^2=k_{J_1}k_{J_2}\ \exp\left[2\left(1+\frac{2}{3}I\right)
-\frac{5}{3}\right]\;;
\eeq
the variant $(b)$ is given by
\begin{eqnarray}
\label{caseb}
{\cal C}_n^{\rm exp\ BLM_b}&=& \frac{x_{J_1}x_{J_2}}{|\vec k_{J_1}|
|\vec k_{J_2}|}\int_{-\infty}^{+\infty}d\nu
\ e^{(Y-Y_0)\left[\bar \alpha_s\left(\mu_R\right)\chi\left(n,\nu\right)
+ \bar \alpha_s^2\left(\mu_R\right)
\left( \bar \chi\left(n,\nu\right)-\frac{T^{\beta}}{C_A}
\chi\left(n,\nu\right)\right)\right]}\alpha_s^2\left(\mu_R\right)
\\
\nonumber
&\times&c_1\left(n,\nu\right) c_2\left(n,\nu\right) \\
&\times& \left[ 1+\alpha_s\left(\mu_R\right)\left(\frac{\beta_0}{4\pi}
\chi\left(n,\nu\right)
- 2\frac{T^{\beta}}{\pi}\right)+\alpha_s\left(\mu_R\right)
\left( \frac{\bar c_1^{\left(1\right)}\left(n,\nu\right)}
{c_1\left(n,\nu\right)}+\frac{\bar c_2^{\left(1\right)}
\left(n,\nu\right)}{c_2\left(n,\nu\right)}\right)
\right]\;, \nonumber
\end{eqnarray}
with $\mu_R$ fixed at the value
\beq
(\mu_R^{\rm BLM})^2=k_{J_1}k_{J_2}\ \exp\left[2\left(1+\frac{2}{3}I\right)
-\frac{5}{3}+\frac{1}{2}\chi\left(\nu,n\right)\right]\;.
\eeq
In Eqs.~(\ref{casea}) and~(\ref{caseb}), we have
\begin{eqnarray*}
T&=&T^{\beta}+T^{conf}\;,\\
T^{\beta}&=&-\frac{\beta_0}{2}\left( 1+\frac{2}{3}I \right)\;,\\
T^{conf}&=& \frac{C_A}{8}\left[ \frac{17}{2}I +\frac{3}{2}\left(I-1\right)\xi
+\left( 1-\frac{1}{3}I\right)\xi^2-\frac{1}{6}\xi^3 \right]\;,
\end{eqnarray*}
where $I=-2\int_0^1dx\frac{\ln\left(x\right)}{x^2-x+1}\simeq 2.3439$ and
$\xi$ is a gauge parameter, fixed at zero, while
$\bar c_i^{(1)}/c_i$ is the NLO impact factor defined as in
Eqs.~(\ref{pure})-(\ref{cis}) with the terms proportional to $\beta_0$
removed.

Results are reported in Tables~\ref{tab:C0-C3_BLM} and~\ref{tab:ratios_BLM}
and in Figs.~\ref{fig:BLM}. We can see that, except for the ratio $C_1/C_0$,
the agreement with experimental data is very good, for both variants, at
the larger values of $Y$.

\section{Discussion}
\label{summary}

In this paper we have studied several, equivalent within the NLA,
representations of the coefficients entering the definition of cross
section, azimuthal decorrelations and ratios of azimuthal decorrelations,
and have compared them with the corresponding CMS experimental data at
the center-of-mass energy of 7 TeV.

We have considered three different procedures to optimize the perturbative
series (PMS, FAC and BLM, the latter in two variants) and found that:

\begin{itemize}

\item the FAC method does not lead to any sensible result for most observables;

\item the ratios $C_2/C_1$ and $C_3/C_2$ are quite well reproduced basically
by all representations treated with the PMS method;

\item the BLM method, implemented in the exponentiated representation,
reproduces quite well all the ratios studied in this work, in the region
$Y\gtrsim 6$; we see, however, a sizeable difference in the theoretical
prediction of the value of $C_0$ between the two variants $(a)$ and $(b)$;
also in Ref.~\cite{Ducloue2014} an important effect on the cross section
is reported when the BLM method is implemented together with an RG-improved
kernel, than with the standard non-RG-improved kernel.

\end{itemize}

\begin{figure}[p]
\centering
  \includegraphics[scale=0.45]{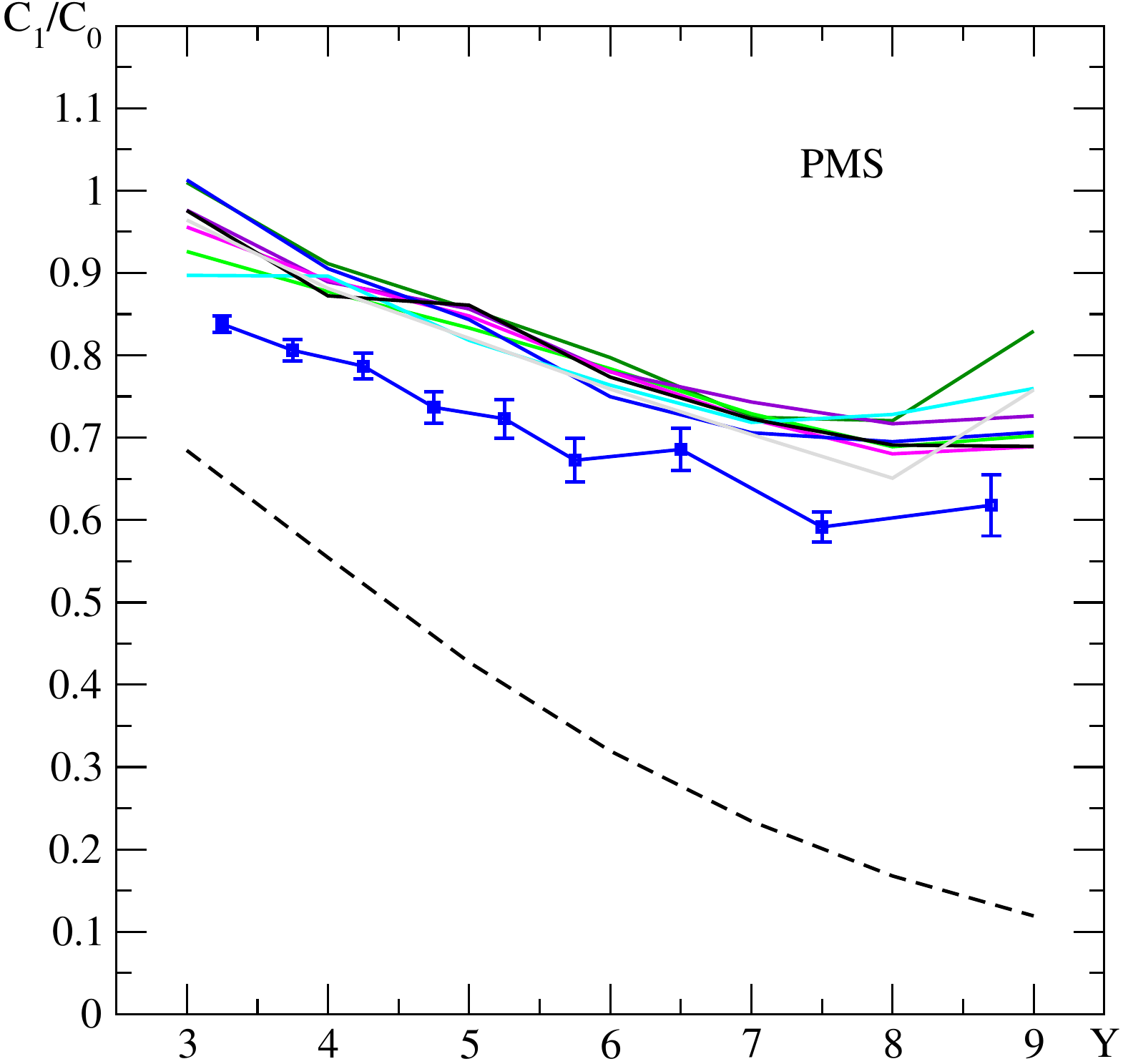}
  \includegraphics[scale=0.45]{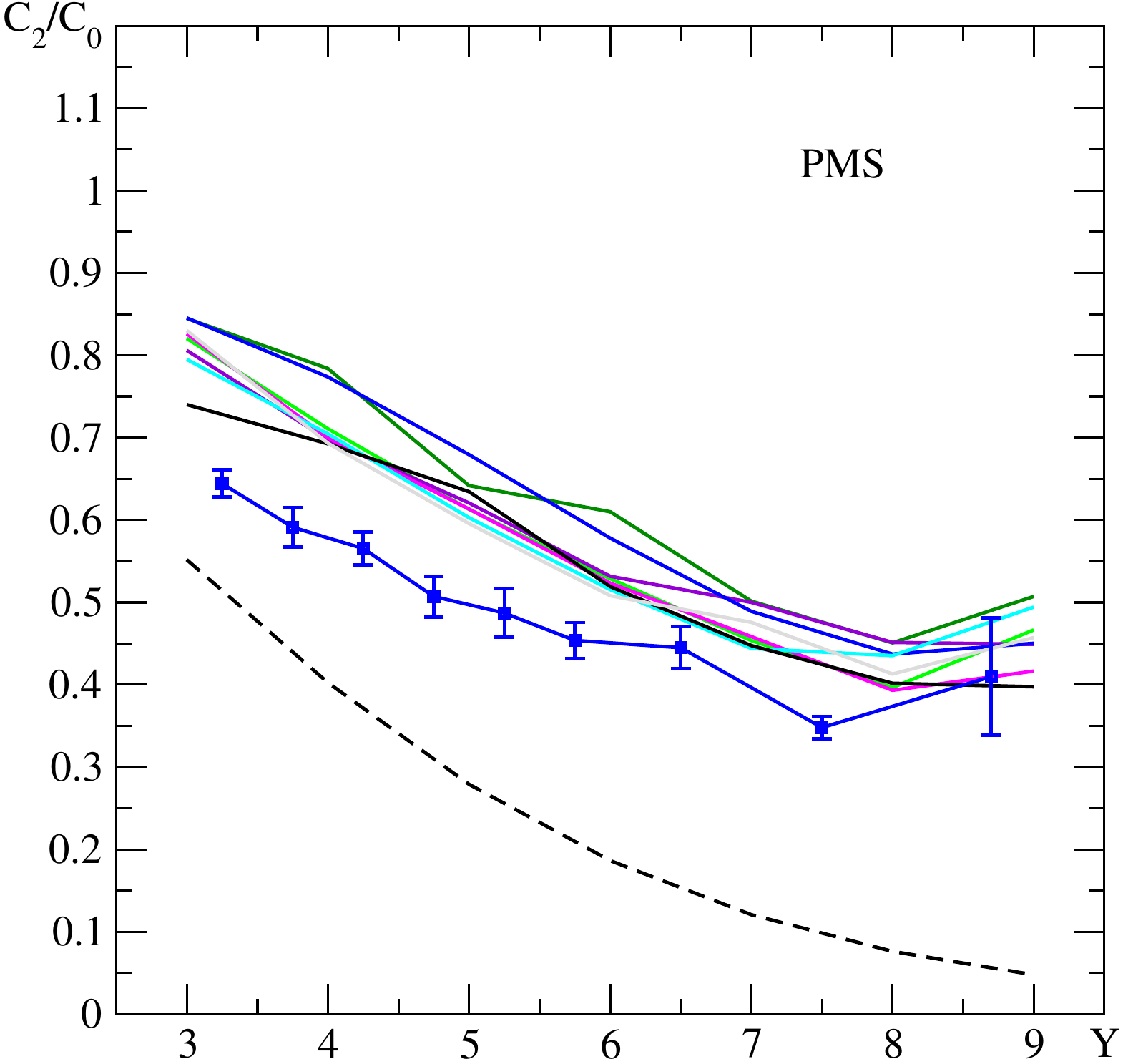}

  \includegraphics[scale=0.45]{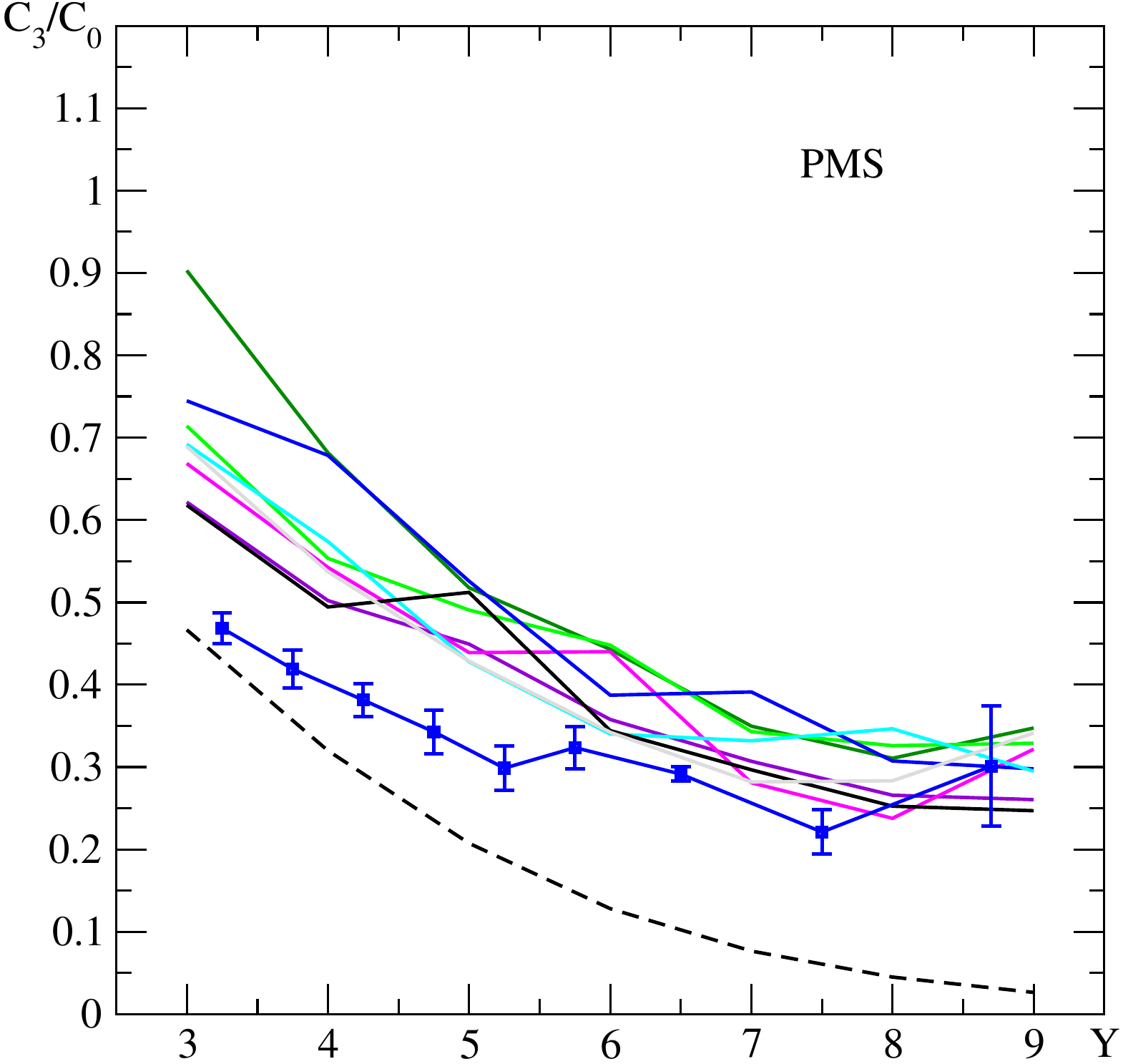}
  \includegraphics[scale=0.45]{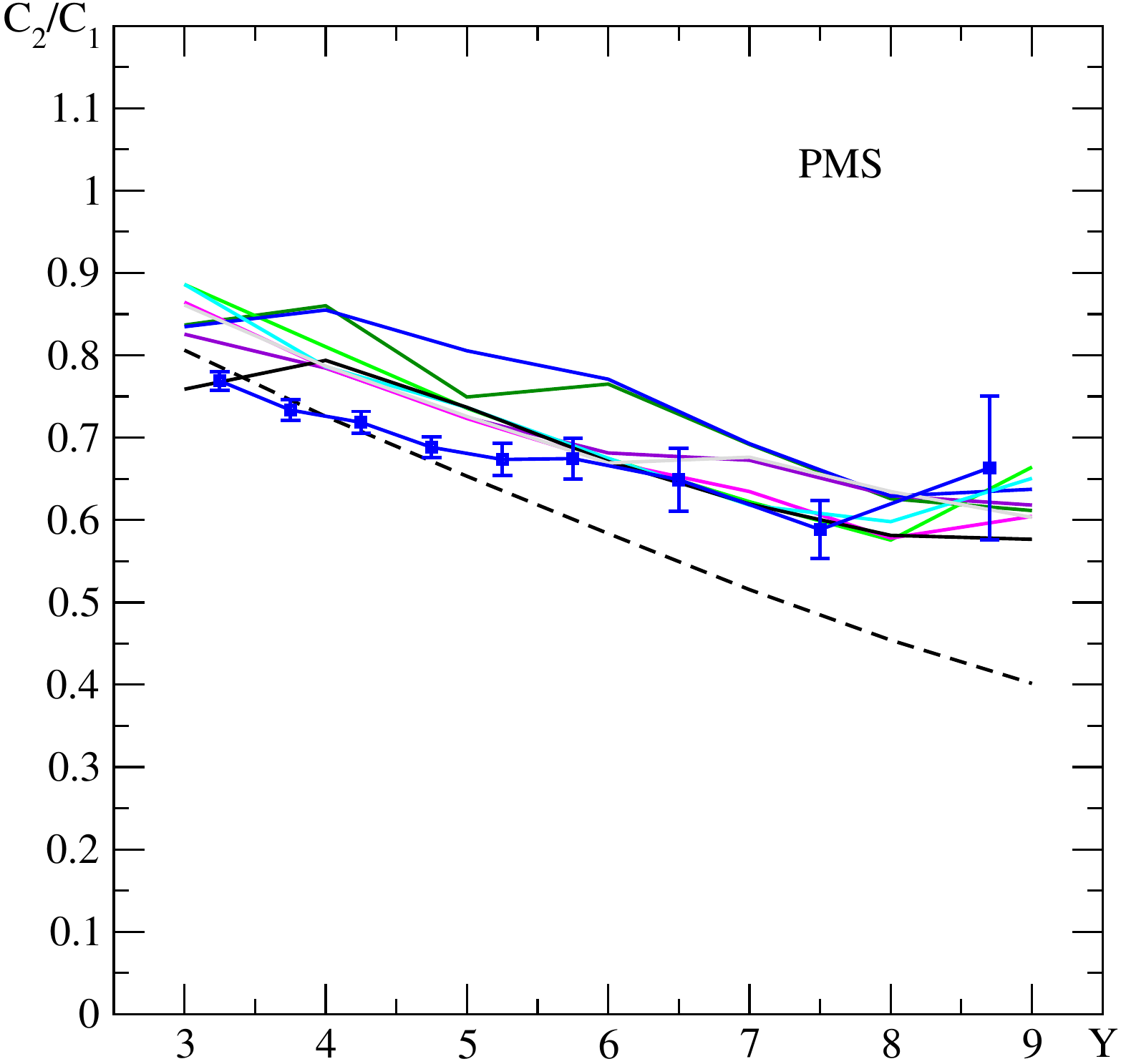}

  \includegraphics[scale=0.45]{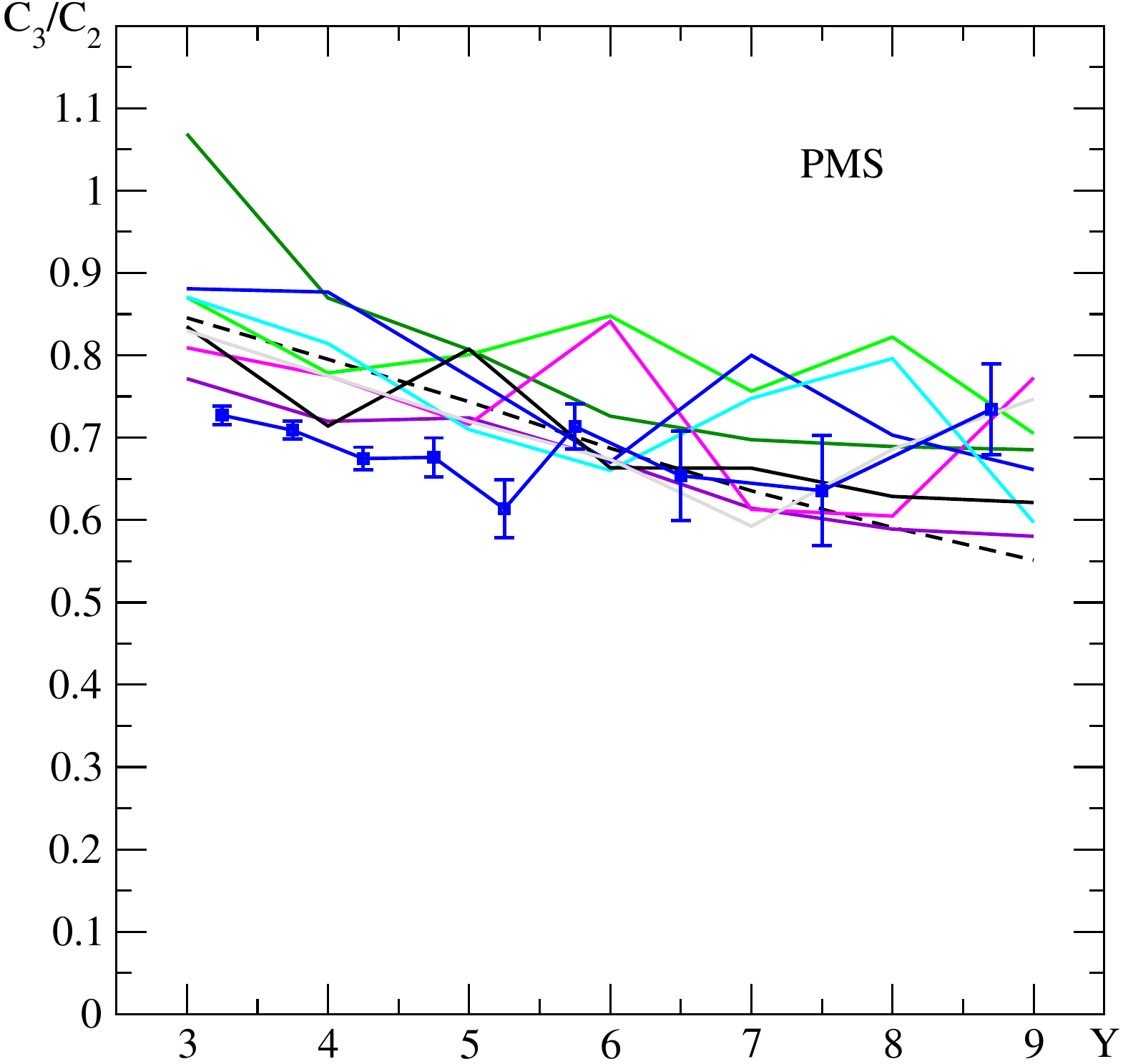}
\caption{$Y$ dependence of several azimuthal correlations and some of their
ratios. Results were obtained with the PMS method. The dashed line
gives the LLA BFKL result; the colored broken lines give the NLA BFKL
result for the eight options NLA$_i$, $i=1,...,8$ considered (see text).}
\label{fig:PMS}
\end{figure}

\begin{figure}[p]
\centering
  \includegraphics[scale=0.42]{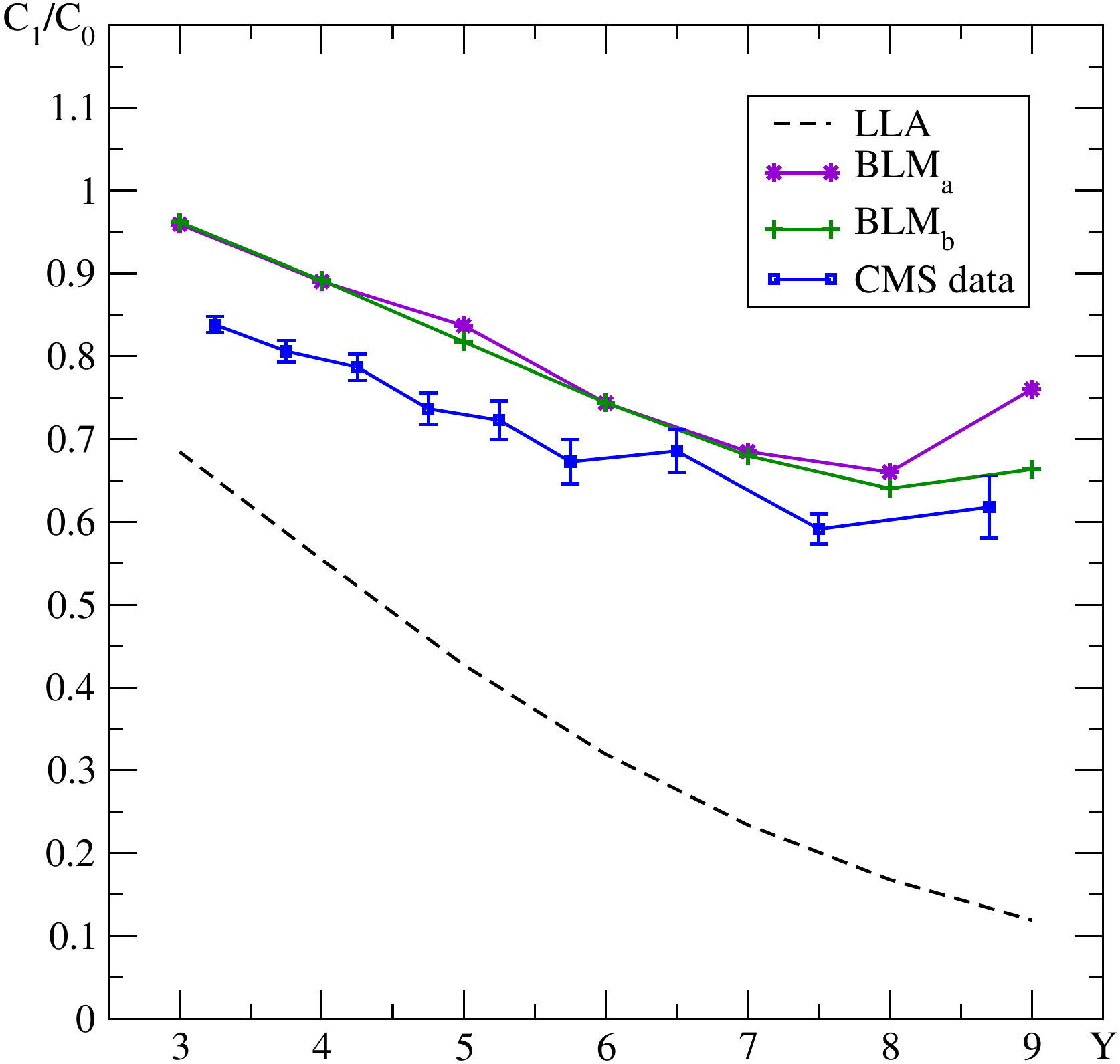}
  \includegraphics[scale=0.42]{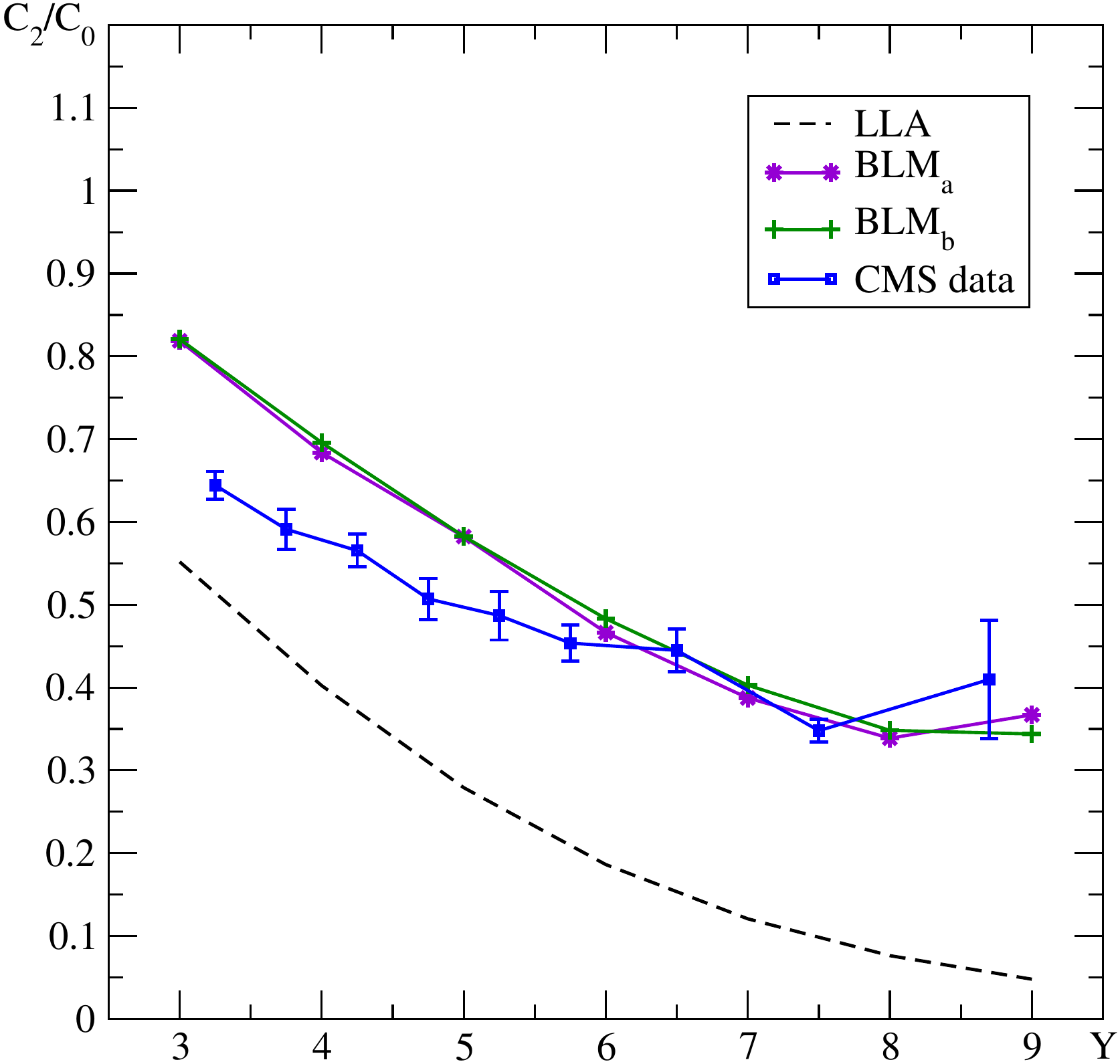}

  \includegraphics[scale=0.42]{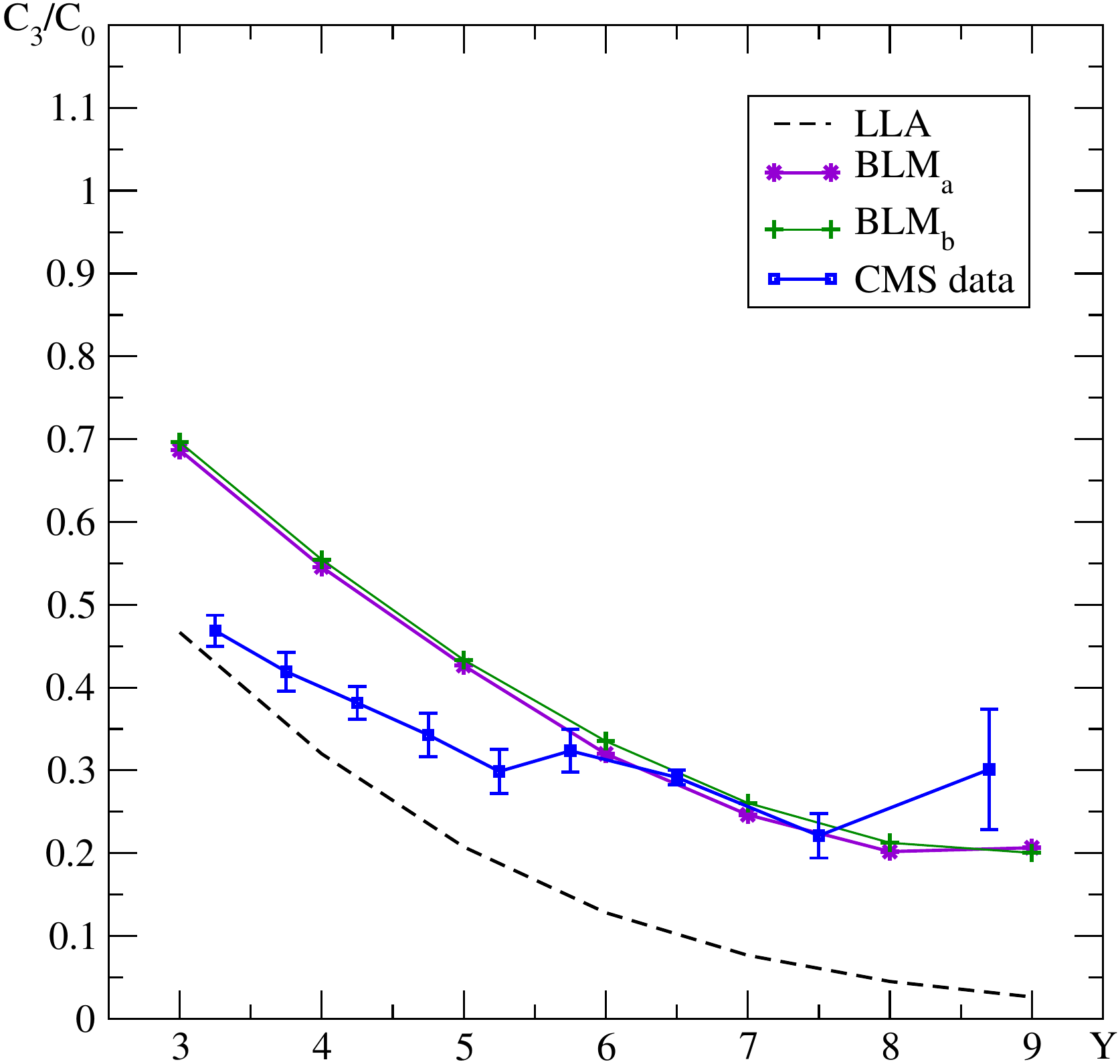}
  \includegraphics[scale=0.42]{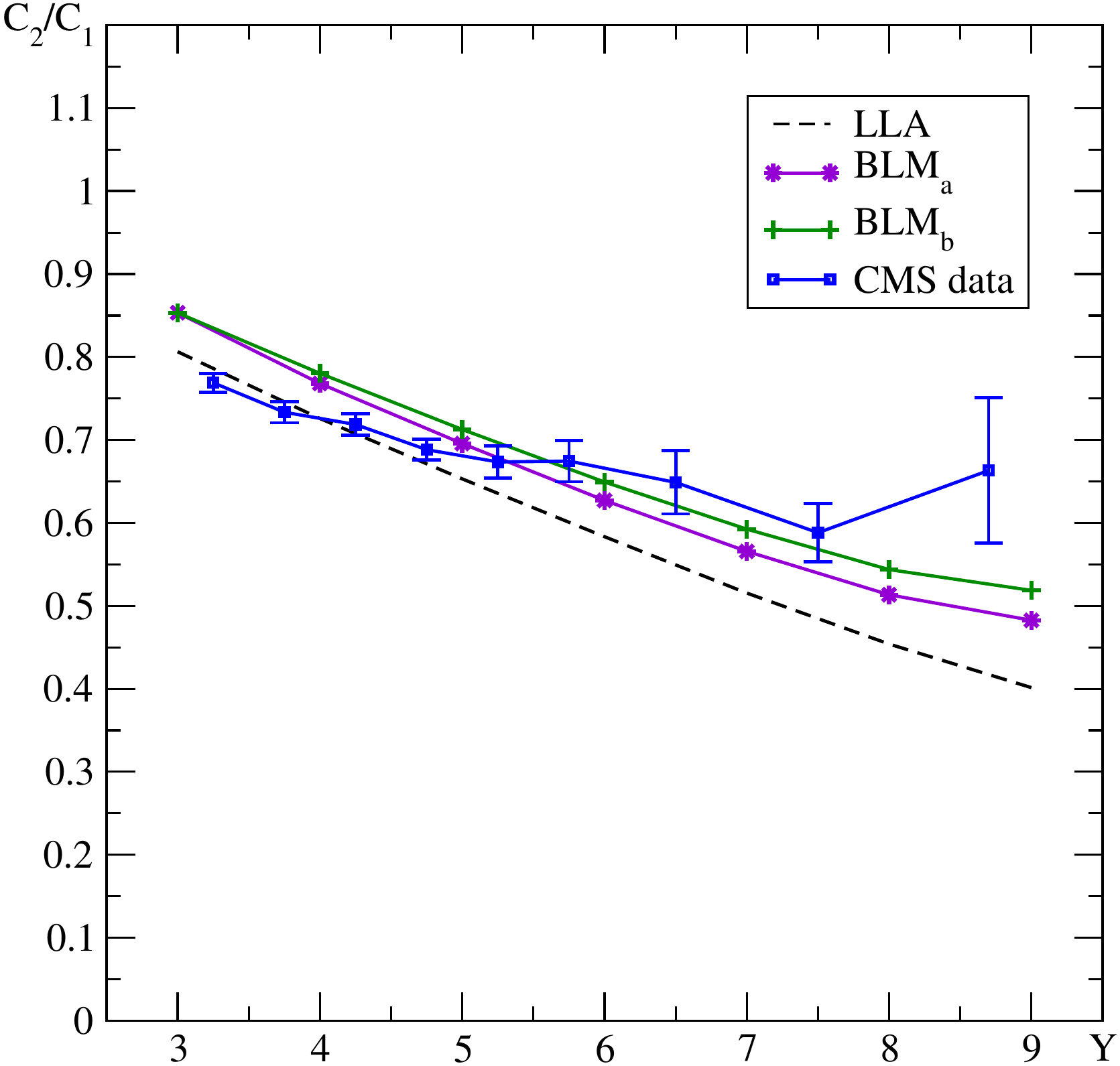}

  \includegraphics[scale=0.42]{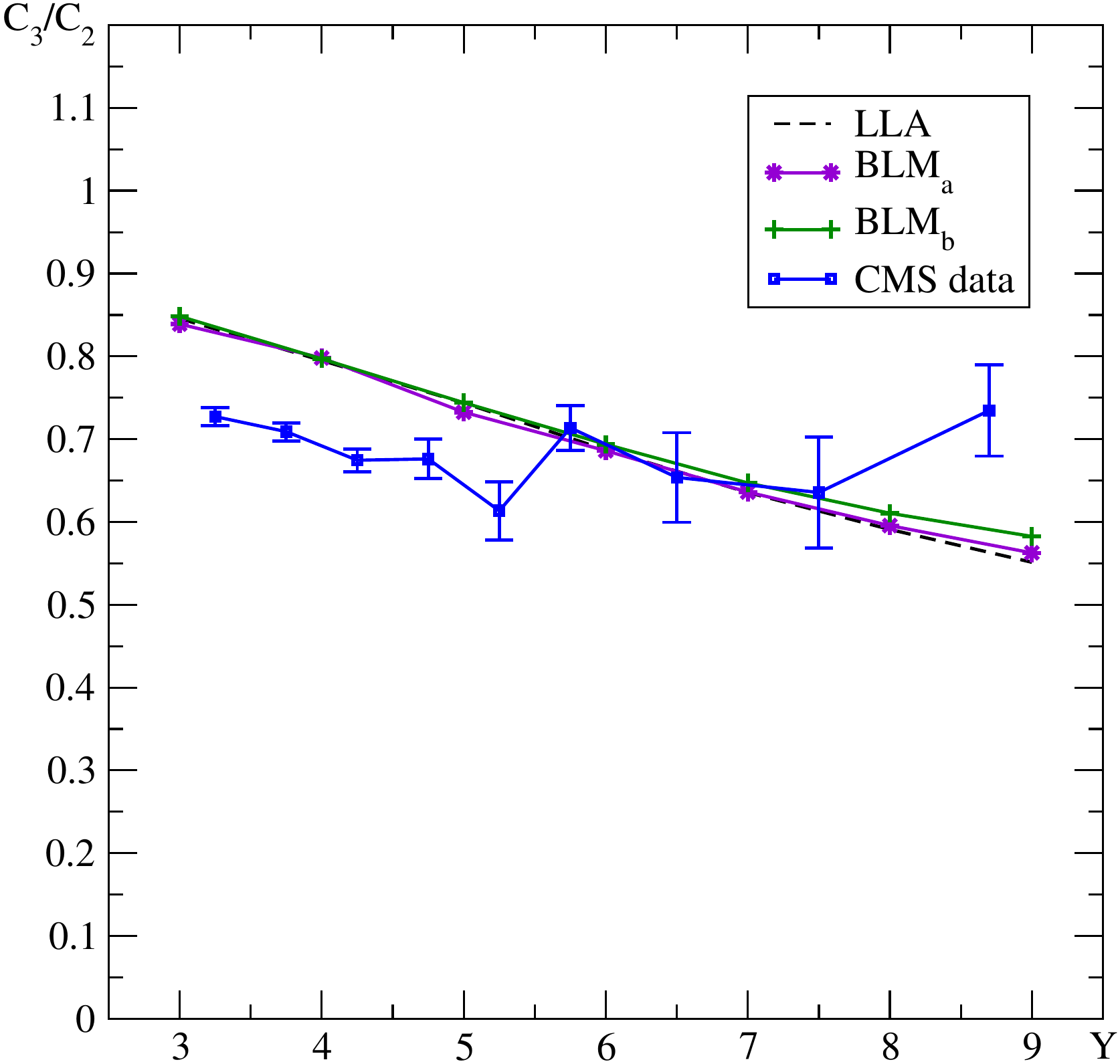}
\caption{$Y$ dependence of several azimuthal correlations and some of their
ratios. Results were obtained with the two variants of the BLM method.
The dashed line gives the LLA BFKL result.}
\label{fig:BLM}
\end{figure}

We believe that the information we gathered in this work can be of
help in preparing new predictions for the same observables considered
the increased collision energy of LHC after the LS1. In particular, it could 
be useful for estimates of theoretical uncertainties. Our numerical analysis 
shows that these uncertainties are rather large, in general due to very large 
NLA BFKL corrections in the considered kinematical range. In particular, the 
plots in Fig.~\ref{fig:PMS} demonstrate that, within the PMS method, results 
obtained using different representations of the NLA BFKL amplitude are quite 
different one from the other. We stress that this type of uncertainty is often 
not considered and in the NLA BFKL analysis one uses just some prescribed 
representation of NLA BFKL amplitude. We believe that one should be aware of
this ``representation'' uncertainty, until the time will come when some 
deeper insight into the physics of effects beyond NLA BFKL will allow to 
choose a definite representation of NLA BFKL amplitude.
Perhaps, the BLM optimization procedure gives us a hint towards the right 
direction, because theoretical predictions derived with BLM~\cite{Ducloue2014} 
turned to be in a rather good agreement with CMS data. Our own 
BLM calculations presented in Fig.~\ref{fig:BLM} support this statement, 
though, comparing our results with the plots of Ref.~\cite{Ducloue2014}, we see
that our predictions lies somewhat beyond the range of the theoretical 
uncertainty bound accepted there. Most probably this difference is related 
with the above mentioned ``representation uncertainty'', indeed our BLM 
amplitudes in Eqs.~(\ref{casea}) and~(\ref{caseb}), in contrast 
with~\cite{Ducloue2014}, do not include the product of the two NLO impact 
factors terms.

Meanwhile, it would be also useful to address, on the experimental side, some
possible issues which could be sources of mismatch with the way
in which Mueller-Navelet jets are defined in theory and that are not
easy to be revealed in the comparison with theoretical predictions,
for being the latter affected in their turn by systematic effects of the same
amount. We list below a few of them.

\begin{itemize}

\item In data analysis defining the $Y$ value for a given final state with two 
jets, the rapidity of one of the two jets could be so small, say 
$|y_{J_i}|\lesssim 2$, that this jet is actually produced in the central 
region, rather than in one of the two forward regions. The longitudinal 
momentum fractions of the parent partons that generate a central jet are very 
small, and one can naturally expect sizable corrections to the vertex of this 
jet, due to the fact that the collinear factorization approach used in the
derivation of the result for jet vertex is not designed for the region of 
small $x$. We believe that a combined theoretical approach that uses collinear 
factorization for the forward and $k_t$-factorization for the central jets 
should be more relevant in such kinematics. 

\item The other issue is related with the experimental event selection for 
Mueller-Navelet jet analysis in a situation when more that two jets are 
detected in one single event. In particular, let us consider events with three 
jets in the final state, two of them being forward in one direction (with 
large positive rapidities, say, $y_1$ and $y_2$ with $y_1>y_2$), and the third 
being forward in the other direction (with large negative rapidity, say, 
$y_3$). Traditionally, as in the current CMS analysis, such event is selected 
as a single Mueller-Navelet jet, where the two selected Mueller-Navelet jets 
are those having the largest interval in rapidity. In our example, these are
the jets with rapidities $y_1$ and $y_3$, so that $Y=y_1-y_3$. This selection 
method is convenient for the experimental analysis, but it does not match the 
definition of Mueller-Navelet jets in the theoretical NLA BFKL calculations.
Examining the derivation of the NLA jet vertex~\cite{IFjet}, one can see that 
what is calculated in the theory is an {\it inclusive} jet production in the  
forward region, with some prescribed values of rapidity and transverse
momentum $\vec k$, where possible additional parton radiation is attributed to 
the inclusive hadron system $X$. Returning to our example of event with three 
detected jets, we see that in order to match the theory it should lead to the
selection of two separate Mueller-Navelet jets events ({\it i.e.} it should
be counted twice): a pair of Mueller-Navelet jets with rapidities $y_1$ and 
$y_3$ (then $Y=y_1-y_3$) and another pair of Mueller-Navelet jets with 
rapidities $y_2$ and $y_3$ (then $Y=y_2-y_3$). This mismatch between 
experimental event selection and theory appears in NLA BFKL and could be 
important due to very large value of NLA BFKL corrections. The issue may be 
clarified either from the experimental side, changing the Mueller-Navelet 
jet selection criterion, or from the theoretical side, which could require 
the generation of separate jet events with Monte Carlo methods.

\item The use of symmetric cuts in the values of $k_{J_i,\rm min}$
maximizes the contribution of the Born term in $C_0$, which is present
for back-to-back jets only and is expected to be large, therefore making
less visible the effect of the BFKL resummation in all observables involving
$C_0$. The use of asymmetric cuts can reduce the contribution of the Born
term and enhance effects with additional undetected hard gluon radiation,
which makes the visibility of BFKL effect more clear in comparison to the 
descriptions based on fixed order DGLAP approach.

\item The experimental determination of the Mueller-Navelet total cross
section, $C_0$, would provide for a yardstick which could help choosing 
a definite NLA representation.

\end{itemize}


\begin{table}[p]
\centering
\caption{$C_1/C_0$ in the LLA and in the NLA according to the eight
different representations discussed in the text; results obtained with
the PMS method.}
\label{tab:C1C0_PMS}
\begin{tabular}{c|ccccccccc}
\hline\noalign{\smallskip}
$Y$ & LLA & NLA$_1$ & NLA$_2$ & NLA$_3$ & NLA$_4$ & NLA$_5$ & NLA$_6$
    & NLA$_7$ & NLA$_8$ \\
\noalign{\smallskip}\hline\noalign{\smallskip}
3 & 0.6845 & 1.0099 & 0.9261 & 0.9762 & 0.9558 & 1.0129 & 0.8970 & 0.9756 & 0.9642 \\
4 & 0.5544 & 0.9112 & 0.8772 & 0.8891 & 0.8915 & 0.9052 & 0.8962 & 0.8721 & 0.8811 \\
5 & 0.4273 & 0.8563 & 0.8332 & 0.8564 & 0.8477 & 0.8433 & 0.8183 & 0.8609 & 0.8207 \\
6 & 0.3195 & 0.7972 & 0.7837 & 0.7799 & 0.7802 & 0.7497 & 0.7637 & 0.7736 & 0.7589 \\
7 & 0.2342 & 0.7248 & 0.7291 & 0.7433 & 0.7224 & 0.7060 & 0.7185 & 0.7226 & 0.7037 \\
8 & 0.1679 & 0.7205 & 0.6889 & 0.7169 & 0.6803 & 0.6951 & 0.7281 & 0.6911 & 0.6508 \\
9 & 0.1192 & 0.8292 & 0.7023 & 0.7262 & 0.6889 & 0.7066 & 0.7596 & 0.6894 & 0.7581 \\
\noalign{\smallskip}\hline
\end{tabular}
\end{table}

\begin{table}[p]
\centering
\caption{$C_2/C_0$ in the LLA and in the NLA according to the eight
different representations discussed in the text; results obtained with
the PMS method.}
\label{tab:C2CO_PMS}
\begin{tabular}{c|ccccccccc}
\hline\noalign{\smallskip}
$Y$ & LLA & NLA$_1$ & NLA$_2$ & NLA$_3$ & NLA$_4$ & NLA$_5$ & NLA$_6$
    & NLA$_7$ & NLA$_8$ \\
\noalign{\smallskip}\hline\noalign{\smallskip}
3 & 0.5519 & 0.8448 & 0.8205 & 0.8059 & 0.8262 & 0.8454 & 0.7949 & 0.7402 & 0.8301 \\
4 & 0.4024 & 0.7838 & 0.7107 & 0.6974 & 0.6997 & 0.7738 & 0.7045 & 0.6924 & 0.6932 \\
5 & 0.2791 & 0.6418 & 0.6130 & 0.6207 & 0.6132 & 0.6794 & 0.6028 & 0.6343 & 0.5956 \\
6 & 0.1864 & 0.6100 & 0.5287 & 0.5316 & 0.5233 & 0.5781 & 0.5154 & 0.5191 & 0.5080 \\
7 & 0.1207 & 0.5014 & 0.4536 & 0.4999 & 0.4584 & 0.4890 & 0.4440 & 0.4475 & 0.4758 \\
8 & 0.0762 & 0.4509 & 0.3966 & 0.4514 & 0.3932 & 0.4372 & 0.4353 & 0.4017 & 0.4130 \\
9 & 0.0479 & 0.5071 & 0.4665 & 0.4489 & 0.4164 & 0.4503 & 0.4942 & 0.3974 & 0.4572 \\
\noalign{\smallskip}\hline
\end{tabular}
\end{table}

\begin{table}[p]
\centering
\caption{$C_3/C_0$ in the LLA and in the NLA according to the eight
different representations discussed in the text; results obtained with
the PMS method.}
\label{tab:C3CO_PMS}
\begin{tabular}{c|ccccccccc}
\hline\noalign{\smallskip}
$Y$ & LLA & NLA$_1$ & NLA$_2$ & NLA$_3$ & NLA$_4$ & NLA$_5$ & NLA$_6$
    & NLA$_7$ & NLA$_8$ \\
\noalign{\smallskip}\hline\noalign{\smallskip}
3 & 0.4667 & 0.9030 & 0.7143 & 0.6218 & 0.6686 & 0.7447 & 0.6920 & 0.6180 & 0.6891 \\
4 & 0.3199 & 0.6816 & 0.5534 & 0.5021 & 0.5424 & 0.6785 & 0.5737 & 0.4944 & 0.5373 \\
5 & 0.2075 & 0.5178 & 0.4907 & 0.4494 & 0.4391 & 0.5258 & 0.4279 & 0.5121 & 0.4286 \\
6 & 0.1281 & 0.4428 & 0.4481 & 0.3577 & 0.4401 & 0.3874 & 0.3403 & 0.3443 & 0.3422 \\
7 & 0.0767 & 0.3497 & 0.3430 & 0.3071 & 0.2810 & 0.3912 & 0.3320 & 0.2966 & 0.2819 \\
8 & 0.0451 & 0.3107 & 0.3260 & 0.2658 & 0.2378 & 0.3073 & 0.3466 & 0.2526 & 0.2833 \\
9 & 0.0264 & 0.3475 & 0.3289 & 0.2605 & 0.3217 & 0.2977 & 0.2951 & 0.2470 & 0.3414 \\
\noalign{\smallskip}\hline
\end{tabular}
\end{table}

\begin{table}[p]
\centering
\caption{$C_2/C_1$ in the LLA and in the NLA according to the eight
different representations discussed in the text; results obtained with
the PMS method.}
\label{tab:C2C1_PMS}
\begin{tabular}{c|ccccccccc}
\hline\noalign{\smallskip}
$Y$ & LLA & NLA$_1$ & NLA$_2$ & NLA$_3$ & NLA$_4$ & NLA$_5$ & NLA$_6$
    & NLA$_7$ & NLA$_8$ \\
\noalign{\smallskip}\hline\noalign{\smallskip}
3 & 0.8063 & 0.8366 & 0.8859 & 0.8256 & 0.8644 & 0.8346 & 0.8862 & 0.7588 & 0.8610 \\
4 & 0.7258 & 0.8602 & 0.8102 & 0.7844 & 0.7849 & 0.8549 & 0.7861 & 0.7939 & 0.7868 \\
5 & 0.6531 & 0.7494 & 0.7357 & 0.7247 & 0.7234 & 0.8055 & 0.7366 & 0.7368 & 0.7257 \\
6 & 0.5835 & 0.7651 & 0.6746 & 0.6815 & 0.6707 & 0.7710 & 0.6749 & 0.6711 & 0.6693 \\
7 & 0.5154 & 0.6917 & 0.6221 & 0.6725 & 0.6346 & 0.6926 & 0.6180 & 0.6193 & 0.6762 \\
8 & 0.4541 & 0.6258 & 0.5757 & 0.6296 & 0.5780 & 0.6290 & 0.5979 & 0.5813 & 0.6346 \\
9 & 0.4015 & 0.6115 & 0.6643 & 0.6181 & 0.6044 & 0.6373 & 0.6506 & 0.5765 & 0.6031 \\
\noalign{\smallskip}\hline
\end{tabular}
\end{table}

\begin{table}[p]
\centering
\caption{$C_3/C_2$ in the LLA and in the NLA according to the eight
different representations discussed in the text; results obtained with
the PMS method.}
\label{tab:C3C2_PMS}
\begin{tabular}{c|ccccccccc}
\hline\noalign{\smallskip}
$Y$ & LLA & NLA$_1$ & NLA$_2$ & NLA$_3$ & NLA$_4$ & NLA$_5$ & NLA$_6$
    & NLA$_7$ & NLA$_8$ \\
\noalign{\smallskip}\hline\noalign{\smallskip}
3 & 0.8456 & 1.0689 & 0.8705 & 0.7716 & 0.8092 & 0.8808 & 0.8706 & 0.8349 & 0.8301 \\
4 & 0.7951 & 0.8696 & 0.7786 & 0.7199 & 0.7752 & 0.8768 & 0.8143 & 0.7141 & 0.7750 \\
5 & 0.7437 & 0.8068 & 0.8006 & 0.7240 & 0.7161 & 0.7740 & 0.7098 & 0.8073 & 0.7196 \\
6 & 0.6871 & 0.7260 & 0.8477 & 0.6729 & 0.8411 & 0.6702 & 0.6602 & 0.6633 & 0.6736 \\
7 & 0.6355 & 0.6975 & 0.7563 & 0.6144 & 0.6130 & 0.7999 & 0.7477 & 0.6629 & 0.5924 \\
8 & 0.5910 & 0.6892 & 0.8221 & 0.5889 & 0.6048 & 0.7030 & 0.7961 & 0.6287 & 0.6859 \\
9 & 0.5513 & 0.6853 & 0.7050 & 0.5803 & 0.7726 & 0.6611 & 0.5971 & 0.6213 & 0.7467 \\
\noalign{\smallskip}\hline
\end{tabular}
\end{table}

\begin{table}[p]
\centering
\caption{$C_0$, $C_1$ and $C_1/C_0$ in the representation NLA$_1$
with the FAC method; columns three, four, six and seven give
the optimal values for $Y_0$ and $\mu_R/\sqrt{|\vec k_{J_1}||\vec k_{J_2}|}$.}
\label{tab:FAC}
\begin{tabular}{c|ccc|ccc|c}
\hline\noalign{\smallskip}
$Y$ & $C_0$ [nb] & $Y_0$ & $n_R$ & $C_1$ [nb] & $Y_0$ & $n_R$ & $C_1/C_0$ \\
\noalign{\smallskip}\hline\noalign{\smallskip}
3 & 2584.9    & 2   & 3.1 & --         & -- & --  & --     \\
4 & 928.303   & 2   & 2.7 & --         & -- & --  & --     \\
5 & 291.053   & 3   & 2.9 & 249.865    & 3  & 2.9 & 0.8585 \\
6 & 75.0725   & 3   & 2.7 & 60.1738    & 3  & 2.3 & 0.8015 \\
7 & 13.8129   & 3.5 & 2.9 & 10.5419    & 3  & 1.9 & 0.7632 \\
8 & 1.23915   & 4   & 3.3 & 0.918584   & 3  & 1.9 & 0.7413 \\
9 & 0.0135918 & 5   & 4.7 & 0.00993695 & 4  & 3.1 & 0.7311 \\
\noalign{\smallskip}\hline
\end{tabular}
\end{table}

\begin{table}[h]
\centering
\caption{$C_0$, $C_1$, $C_2$ and $C_3$ in the representation NLA$_1$
with the BLM method, in both variants $(a)$ and $(b)$.}
\label{tab:C0-C3_BLM}
\begin{tabular}{c|cc|cc|cc|cc}
\hline\noalign{\smallskip}
    & \multicolumn{2}{c|}{$C_0$ [nb]} & \multicolumn{2}{c|}{$C_1$ [nb]}
    & \multicolumn{2}{c|}{$C_2$ [nb]} & \multicolumn{2}{c}{$C_3$ [nb]} \\
$Y$ & $(a)$ & $(b)$ & $(a)$ & $(b)$ & $(a)$ & $(b)$ & $(a)$ & $(b)$ \\
\noalign{\smallskip}\hline\noalign{\smallskip}
3 & 2240.88   & 2267.12   & 2150.07    & 2180.83 & 1834.76 & 1860.73 & 1539.24 & 1578.51 \\
4 & 794.934   & 814.956   & 707.761    & 726.72 & 543.584 & 567.018 & 433.784 & 451.898 \\
5 & 237.577   & 252.277   & 198.863    & 206.239 & 138.340 & 146.97 & 101.337 & 109.352 \\
6 & 61.6366   & 64.3728   & 45.8401    & 47.901 & 28.7511 & 31.1006 & 19.7234 & 21.5774 \\
7 & 11.1072   & 11.7626   & 7.60735    & 7.99795 & 4.3031  & 4.73926 & 2.73730 & 3.0664 \\
8 & 0.96651  & 1.05596   & 0.63085   & 0.67637 & 0.32757 & 0.36776 & 0.19508 & 0.22453 \\
9 & 0.00911 & 0.01119 & 0.00693 & 0.00742 & 0.00334 & 0.00385 & 0.00188 &0.00224 \\
\noalign{\smallskip}\hline
\end{tabular}
\end{table}

\begin{table}[h]
\centering
\caption{$C_1/C_0$, $C_2/C_0$, $C_3/C_0$, $C_2/C_1$, $C_3/C_2$
in the representation NLA$_1$ with the BLM method,
in both variants $(a)$ and $(b)$.}
\label{tab:ratios_BLM}
\begin{tabular}{c|cc|cc|cc|cc|cc}
\hline\noalign{\smallskip}
    & \multicolumn{2}{c|}{$C_1/C_0$} & \multicolumn{2}{c|}{$C_2/C_0$}
    & \multicolumn{2}{c|}{$C_3/C_0$} & \multicolumn{2}{c|}{$C_2/C_1$}
    & \multicolumn{2}{c}{$C_3/C_2$} \\
$Y$ & $(a)$ & $(b)$ & $(a)$ & $(b)$ & $(a)$ & $(b)$ & $(a)$ & $(b)$
    & $(a)$ & $(b)$ \\
\noalign{\smallskip}\hline\noalign{\smallskip}
3 & 0.9595 & 0.9619 & 0.8188 & 0.8207 & 0.6869 & 0.6963 & 0.8533 & 0.8532 & 0.8389 & 0.8483 \\
4 & 0.8903 & 0.8917 & 0.6838 & 0.6958 & 0.5457 & 0.5545 & 0.7680 & 0.7802 & 0.7980 & 0.7970 \\
5 & 0.8370 & 0.8175 & 0.5823 & 0.5826 & 0.4265 & 0.4335 & 0.6957 & 0.7126 & 0.7325 & 0.7440 \\
6 & 0.7437 & 0.7441 & 0.4465 & 0.4831 & 0.3200 & 0.3352 & 0.6272 & 0.6493 & 0.6860 & 0.6938 \\
7 & 0.6849 & 0.6799 & 0.3874 & 0.4029 & 0.2464 & 0.2607 & 0.5657 & 0.5926 & 0.6361 & 0.6470 \\
8 & 0.6602 & 0.6405 & 0.3389 & 0.3483 & 0.2018 & 0.2126 & 0.5134 & 0.5437 & 0.5955 & 0.6105 \\
9 & 0.7604 & 0.6634 & 0.3670 & 0.3441 & 0.2065 & 0.2005 & 0.4826 & 0.5187 & 0.5627 & 0.5826 \\
\noalign{\smallskip}\hline
\end{tabular}
\end{table}

\clearpage


\vspace{1.0cm} \noindent
{\Large \bf Acknowledgments} \vspace{0.5cm}

D.I. thanks the Dipartimento di Fisica dell'Universit\`a della Calabria and
the Isti\-tu\-to Na\-zio\-na\-le di Fisica Nucleare (INFN), Gruppo collegato 
di Cosenza, for warm hospitality and financial support. The work of D.I. was 
also supported in part by the grant RFBR-13-02-00695-a.
\\
F.C. and B.M thank A.~Sabio Vera for fruitful discussions and the 
In\-sti\-tu\-to de F\'{\i}\-si\-ca Te\'o\-ri\-ca/UAM-CSIC, for warm 
hospitality during the early stages of this work.
\\
The work of B.M. was supported in part by the grant RFBR-13-02-90907 and by 
the European Commission, European Social
Fund and Calabria Region, that disclaim any liability for the use that can be
done of the information provided in this paper.
\\
We thank F.G.~Celiberto for noticing a mismatch between the content of
Tables~7 and~8 of the previous version of the paper and the plots in Fig.~3 
and for checking all entries in these Tables by an independent 
\textsc{Fortran} program.


\begin{thebibliography}{}
%
%

\bibitem{Mueller:1986ey}
A.H.~Mueller and H.~Navelet, Nucl. Phys. B282, 727 (1987).

\bibitem{BFKL}
V.S.~Fadin, E.A.~Kuraev, L.N.~Lipatov, Phys. Lett. B60, 50 (1975);
E.A.~Kuraev, L.N.~Lipatov and V.S.~Fadin, Zh. Eksp. Teor. Fiz. 71, 840 (1976)
[Sov. Phys. JETP 44, 443 (1976)]; 72, 377 (1977) [45, 199 (1977)];
Ya.Ya.~Balitskii and L.N.~Lipatov, Sov. J. Nucl. Phys. 28, 822 (1978).

\bibitem{DGLAP}
V.N.~Gribov, L.N.~Lipatov, Sov. J. Nucl. Phys. 15, 438 (1972);
G.~Altarelli, G.~Parisi, Nucl. Phys. B126, 298 (1977);
Y.L.~Dokshitzer, Sov. Phys. JETP 46, 641 (1977).

\bibitem{sabioV}
A.~Sabio Vera and F.~Schwennsen, Nucl.Phys. B 776 (2007) 170;
A.~Sabio Vera, Nucl.Phys. B 746 (2006) 1.

\bibitem{NLA-kernel}
V.S.~Fadin and L.N.~Lipatov, Phys. Lett. B 429 (1998) 127;
M.~Ciafaloni and G.~Camici, Phys. Lett. B 430 (1998) 349.

\bibitem{Ivanov2006}
D.Yu.~Ivanov and A.~Papa, Nucl. Phys. B 732, 183 (2006);
Eur. Phys. J. C 49, 947 (2007);
F.~Caporale, A.~Papa and A.~Sabio Vera, Eur. Phys. J. C 53, 525 (2008).

\bibitem{IFjet}
J.~Bartels, D.~Colferai and G.P.~Vacca, Eur. Phys. J. C24, 83-99 (2002);
Eur. Phys. J. C29, 235-249 (2003);
F.~Caporale, D.Yu~Ivanov, B.~Murdaca, A.~Papa and A.~Perri, JHEP 1202, 101
(2012).

\bibitem{SCA}
D.Yu~Ivanov and A.~Papa, JHEP 1205, 086 (2012).

\bibitem{collinear}
G.P.~Salam, JHEP 9807 (1998) 019;
M.~Ciafaloni, D.~Colferai, G.P.~Salam, A.M.~Stasto, Phys. Lett. B 587 (2004)
87; Phys. Rev. D 68 (2003) 114003; Phys. Lett. B 576 (2003) 143;
Phys. Lett. B 541 (2002) 314; Phys. Rev. D 66 (2002) 054014;
M.~Ciafaloni, D.~Colferai, G.P.~Salam, JHEP 0007 (2000) 054;
JHEP 9910 (1999) 017; Phys. Rev. D 60 (1999) 114036;
M.~Ciafaloni, D.~Colferai, Phys. Lett. B 452 (1999) 372;
A.~Sabio Vera, Nucl. Phys. B 722 (2005) 65.

\bibitem{PMS}
P.M.~Stevenson, Phys. Lett. B100, 61 (1981); Phys. Rev. D 23, 2916 (1981).

\bibitem{FAC}
G.~Grunberg, Phys. Lett. B95, 70 (1980) [Erratum-ibid. B110, 501 (1982)];
ibid. B114, 271 (1982); Phys. Rev. D29, 2315 (1984).

\bibitem{BLM}
S.J.~Brodsky, G.P.~Lepage, P.B.~Mackenzie, Phys. Rev. D 28, 228 (1983).

\bibitem{Colferai2010}
D.~Colferai, F.~Schwennsen, L.~Szymanowski, S.~Wallon, JHEP 1012, 026 (2010).

\bibitem{Caporale2013}
F.~Caporale, D.Yu~Ivanov, B.~Murdaca and A.~Papa, Nucl. Phys. B 877, 73 (2013).



\bibitem{Salas2013}
F.~Caporale, B.~Murdaca, A.~Sabio Vera and C.~Salas, Nucl. Phys. B875,
134 (2013).

\bibitem{CMS}
CMS Collaboration, S.~Chatrchyan {\it et al.}, CMS PAS FSQ-12-002.

\bibitem{Ducloue2013}
B.~Duclou\'e, L.~Szymanowski, S.~Wallon, JHEP 1305, 096 (2013).

\bibitem{Ducloue2014}
B.~Duclou\'e, L.~Szymanowski, S.~Wallon, Phys. Rev. Lett. 112, 082003 (2014).

\bibitem{Ducloue:2014koa}
 B.~Duclou\'e, L.~Szymanowski, S.~Wallon,
  arXiv:1407.6593 [hep-ph].

\bibitem{PDF}
A.D.~Martin, W.J.~Stirling, R.S.~Thorne and G.~Watt, Eur. Phys. J. C 63, 189
(2009).

\bibitem{BLMpaper}
F.~Caporale, D.Yu.~Ivanov, B.~Murdaca and A.~Papa, {\emph{in preparation}}.

\end{thebibliography}


\end{document}